\documentclass[a4paper,english]{iopart}
\usepackage[T1]{fontenc}
\usepackage{graphicx}

\makeatletter

\usepackage{iopams}
\usepackage{setstack}

\usepackage{babel}
\makeatother

\begin{document}

\title{Charge Excitations in Cuprate and Nickelate in Resonant Inelastic
X-Ray Scattering}

\author{Manabu Takahashi,$^{1}$ Junichi Igarashi,$^{2}$ and Taeko Semba$^{2}$}

\address{$^{1}$Faculty of Engineering, Gunma University, Kiryu, Gunma 376-8515,
Japan\\
$^{2}$Faculty of Science, Ibaraki University, Mito, Ibaraki 310-8512,
Japan}

\begin{abstract}
We analyze the resonant inelastic x-ray scattering (RIXS) spectra
at the Cu and Ni $K$ edges in $\mathrm{La_{2}CuO_{4}}$ and $\mathrm{La_{2}NiO_{4}}$,
respectively. We make use of the Keldysh-Green-function formalism,
in which the RIXS intensity is described by a product of incident-photon-dependent
factor and density-density correlation function in the $3d$ states.
The former factor is calculated using the $4p$ density of states
given by an ab initio band structure calculation and the latter using
the wavefunctions given by a Hartree-Fock calculation of a multi-orbital
tight-binding model. The initial state is described within the Hartree-Fock
approximation and the electron correlations on charge excitations
are treated within the random phase approximation. The calculated
RIXS spectra well reproduce several characteristic features in the
experiments. Although several groups have interpreted the RIXS peaks
as bound excitons, our calculation indicates that they should be interpreted
as band-to-band excitations augmented by electron correlations. The
difference in RIXS spectra between $\mathrm{La_{2}CuO_{4}}$ and $\mathrm{La_{2}NiO_{4}}$
is explained from this point of view. 
\end{abstract}



\maketitle

\section{Introduction}

Investigation of the electronic excitations in the transition metal
compounds is fundamental for understanding their electronic properties.
The excitations may be characterized into two types, spin and charge
excitations. For the former, the inelastic neutron scattering technique
is quite useful to investigate energy-momentum relations. For the
latter, taking advantage of strong synchrotron sources, the resonant
inelastic x-ray scattering (RIXS) technique has become a powerful
tool to probe charge excitations and to investigate their energy-momentum
relations. For the RIXS tuned at transition metal $K$-edge, the process
is described as a second-order optical process, in which a $1s$-core
electron is excited to an empty $4p$ state by absorbing a photon,
then charge excitations are created in the $3d$ states to screen
the core-hole potential, and finally the photoexcited $4p$ electron
recombines with the $1s$-core hole by emitting a photon. In the end,
the charge excitations are left behind.

RIXS measurement at the $\mathrm{Cu}$ $K$-edge has been carried
out to search for charge excitation modes and to determine the momentum
dependence of charge excitation in high $T_{\mathrm{C}}$ cuprates
and their related materials for better understanding of the unconventional
high-$T_{\mathrm{C}}$ superconductivity. For the undoped material
$\mathrm{La_{2}CuO_{4}}$, the RIXS spectra as a function of energy
loss are composed of several peaks and show clear momentum dependence\cite{Kim2002,Lu2006,Ellis2008PRB77,Collart2006}.
The peak position in terms of energy loss shifts remarkably as the
momentum transfer changes. Several groups have claimed that the RIXS
intensity reflects a bound exciton and interpreted the peak-shifts
as a dispersion relation of bound exciton\cite{Collart2006,Zhang1998PRB58}.
The RIXS experiment at $\mathrm{Ni}$ $K$-edge on $\mathrm{La_{2}NiO_{4}}$,
which has isostructure with $\mathrm{La_{2}CuO_{4}}$, has been also
carried out in order to clarify the difference of charge excitations
from $\mathrm{La_{2}CuO_{4}}$. In contrast to the cuprate, no superconductivity
has been reported in the nickelate. It has been reported that the
RIXS spectra are composed of several peaks which do not show clear
momentum dependence. The peak position in terms of energy loss does
not shift even though the momentum transfer varies. It has been suggested
that the exciton is well localized in the nickelate, while it is mobile
in the cuprate\cite{Collart2006}. 

On the other hand, it has been argued that the RIXS intensity should
be interpreted as band-to-band excitations augmented by electron correlations,
not as bound excitons. Nomura and Igarashi (NI)\cite{Nomura2004,Nomura2005}
have proposed a general formalism of the RIXS spectra by extending
the resonant Raman theory developed by Nozi\'eres and Abrahams\cite{Nozieres1974}
on the basis of the many-body formalism of Keldysh. In the NI formula,
with the assistance of the Born approximation on the interaction between
the $1s$ hole and $3d$ electrons, the RIXS intensity is described
by a product of an incident-photon-dependent factor and a density-density
correlation function in the $3d$ states. The NI formula has been
successfully applied to the quasi-one-dimensional cuprates $\mathrm{SrCuO_{3}}$\cite{Nomura2004},
$\mathrm{CuGeO_{3}}$\cite{Suga2005}, two-dimensional cuprate $\mathrm{La_{2}CuO_{4}}$\cite{Nomura2005,Igarashi2006,Takahashi2008JPSJ77},
and the prototypical AFM insulator $\mathrm{NiO}$\cite{Takahashi2007NiO}.
In these studies, the electronic structures in the AFM phase have
been calculated within the Hartree-Fock approximation (HFA) using
tight-binding models. It is known that the HFA works well for describing
the electronic structures in the AFM insulators. Two-particle correlations
have been taken into account within the random phase approximation
(RPA), which has been found to modify the spectral shape as a function
of energy loss, having led to a good agreement with the experiments.
On the basis of these successes, we may conclude that the RIXS intensity
arises from band-to-band transitions to screen the core-hole potential
in the intermediate state. Multiple-scattering contributions due to
the core-hole potential have been also investigated in order to examine
the validity of the Born approximation, because the core-hole potential
is not definitely weak. Having evaluated the contributions by means
of the time-representation method by Nozi\`eres and De Dominicis\cite{Nozieres1969_1097},
it was found that the contributions could be mainly absorbed into
the shift of the core-level energy with minor modifications of the
RIXS spectral shape\cite{Igarashi2006}. This result partly justifies
the use of the Born approximation. 

In the present study, we discuss the RIXS intensity and clarify the
origin of the difference of them in $\mathrm{La_{2}CuO_{4}}$ and
$\mathrm{La_{2}NiO_{4}}$ by applying the NI formula. We exploit the
tight-binding HFA calculation and the ab initio band structure calculation
based on the local density approximation(LDA) to calculate the density-density
correlation function and the incident photon dependent factor, respectively.
We obtain stable AFM insulating solution having an energy gap about$1.7$($3.5$)
eV within the HFA for $\mathrm{La_{2}CuO_{4}}$ ($\mathrm{La_{2}NiO_{4}}$).
Note that the band structure calculation with the LDA fails to describe
the AFM insulating state for these systems. The two-particle correlations
in the intermediate states are treated within the RPA. We obtain the
RIXS intensities in a range of energy loss $2\sim6$ ($4\sim8$) eV
for $\mathrm{La_{2}CuO_{4}}$ ($\mathrm{La_{2}NiO_{4}}$). The calculated
RIXS spectra well reproduce several characteristic features in the
experiments\cite{Lu2006,Ellis2008PRB77,Collart2006}. Particularly,
the calculated RIXS spectra for $\mathrm{La_{2}CuO_{4}}$ exhibit
the noticeable momentum dependence of the peak structure while those
for $\mathrm{La_{2}NiO_{4}}$ do only slightly, being consistent with
the experiments. Although our results do not reproduce all features
in the RIXS spectra at this stage partly because of the strong correlation
between $3d$ electrons, our results suggest that the RIXS intensity
should be interpreted as band-to-band transition and the peak shifts
as a function of momentum should not be interpreted as a dispersion
relation of a kind of exciton, because the peaks with broad width
constitute an energy continuum generated by band-to-band transition.

The present paper is organized as follows. In Sec. 2, we briefly summarize
the NI formula for the RIXS spectra. In Sec.3 we discuss the calculated
RIXS spectra comparing with the experiments in connection with the
electronic structure within the HFA in the AFM phase of $\mathrm{La_{2}CuO_{4}}$
and $\mathrm{La_{2}NiO_{4}}$. The last section is devoted to the
concluding remarks.

\section{RIXS Formula}

\begin{figure}
\begin{centering}
\includegraphics[clip,scale=0.5]{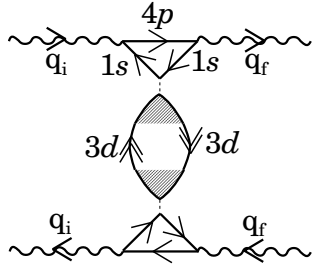} 
\par\end{centering}

\caption{\label{fig:diagram1}Diagrammatic representation for the RIXS intensity.
The wavy and solid lines represent photon and electron bare Green's
functions, respectively. The dotted line is the core-hole interaction
$V$. The solid lines with double arrows are the Keldysh type Green's
functions. The shaded area represents the effective scattering vertex
renormalized by $3d$-$3d$ Coulomb interaction in the RPA. }

\end{figure}

We briefly summarize the NI formula for the RIXS. In the RIXS process,
an incident photon is absorbed by exciting a $\mathrm{Cu}$($\mathrm{Ni}$)
$1s$ core electron to the unoccupied $\mathrm{Cu}$($\mathrm{Ni}$)
$4p$ state, and a photon is emitted by recombining the $4p$ electron
and the core hole. In the intermediate state of the RIXS process,
the core-hole potential acts on the $3d$ states and creates electron-hole
pairs in order to screen the core-hole potential. In the end of the
process, single electron-hole pair is left behind within the Born
approximation, carrying momentum-energy $q\equiv\left(\mathbf{q},\omega\right)=\left(\mathbf{q}_{i}-\mathbf{q}_{f},\omega_{i}-\omega_{f}\right)$,
where $q_{i}=\left(\mathbf{q}_{i},\omega_{i}\right)$ and $q_{f}=\left(\mathbf{q}_{f},\omega_{f}\right)$
are momentum-energy of incident and scattered photons, respectively.
The RIXS intensity is derived on the basis of the Keldysh-Green function
scheme. The process is diagrammatically shown in Fig.\ref{fig:diagram1}.
Within the Born approximation to the core-hole potential, the RIXS
intensity is given by, \begin{eqnarray}
W(q_{i},\mathbf{e}_{i};q_{f},\mathbf{e}_{f}) & = & \frac{N\left|w\right|^{4}}{4\omega_{i}\omega_{f}}\sum_{\lambda m\sigma}\sum_{\lambda'm'\sigma'}\nonumber \\
 &  & \times Y_{\lambda m\sigma,\lambda'm'\sigma'}^{+-}(q)J_{\mathrm{B}\lambda\lambda'}\left(\omega_{i},\mathbf{e}_{i};\omega_{f},\mathbf{e}_{f}\right),\label{eq:general}\end{eqnarray}
 where $\lambda$ indicates the $\mathrm{Cu}$($\mathrm{Ni}$) site
$\lambda$ in a unit cell. $N$ is the number of unit cell and $w$
is the $1s$-$4p$ dipole transition matrix element, which is assumed
to be constant. The factor $J_{\mathrm{B}\lambda\lambda'}\left(\omega_{i},\mathbf{e}_{i};\omega_{f},\mathbf{e}_{f}\right)$
describes the incident-photon dependence, which is given by \begin{eqnarray}
J_{\mathrm{B}\lambda\lambda'}\left(\omega_{i},\mathbf{e}_{i};\omega_{f},\mathbf{e}_{f}\right) & = & \left(\sum_{\eta\eta'}e_{i\eta}L_{\mathrm{B}\lambda}^{\eta\eta'}\left(\omega_{i};\omega_{f}\right)e_{f\eta'}\right)\nonumber \\
 &  & \times\left(\sum_{\eta\eta'}e_{i\eta}L_{\mathrm{B}\lambda'}^{\eta\eta'}\left(\omega_{i};\omega_{f}\right)e_{f\eta'}\right)^{*},\label{eq:IPEF}\end{eqnarray}
 where $e_{i\eta}$ ($e_{f\eta}$) is the $\eta$ component of the
polarization vector $\mathbf{e}_{i}$ ($\mathbf{e}_{f}$) with $\eta=x,y,z$,
and $L_{\mathrm{B}\lambda}^{\eta\eta'}\left(\omega_{i};\omega_{f}\right)$
is given by

\begin{eqnarray}
L_{\mathrm{B}\lambda}^{\eta\eta'}\left(\omega_{i};\omega_{f}\right) & = & \frac{V}{N}\int_{\epsilon_{0}}^{\infty}\frac{\rho_{\lambda4p}^{\eta\eta'}\left(\epsilon\right){\rm d}\epsilon}{\left(\omega_{i}+\epsilon_{1s}+i\Gamma_{1s}-\epsilon\right)\left(\omega_{f}+\epsilon_{1s}+i\Gamma_{1s}-\epsilon\right)}.\label{eq.born}\end{eqnarray}
 $V$ represents the $1s$ core-hole potential acting on the $3d$
electrons, $\Gamma_{1s}$ is the life-time broadening width of the
core-hole state, and the lower limit of the integral $\epsilon_{0}$
indicates the energy at the bottom of the conduction band. This expression
comes from the upper triangle in Fig.\ref{fig:diagram1}. The $\rho_{\lambda4p}^{\eta\eta'}$
is the DOS matrix in the $p$ symmetric states at the $\mathrm{Cu}$($\mathrm{Ni}$)
site $\lambda$, which may be given by \begin{equation}
\rho_{\lambda4p}^{\eta\eta'}\left(\epsilon\right)=\sum_{\sigma}\sum_{n\mathbf{k}}\phi_{\lambda\eta\sigma}^{*}\left(n,\mathbf{k}\right)\phi_{\lambda\eta'\sigma}\left(n,\mathbf{k}\right)\delta\left(\epsilon-\epsilon_{n}\left(\mathbf{k}\right)\right),\label{eq:4p_dos_mat}\end{equation}
 where $\phi_{\lambda\eta\sigma}\left(n,\mathbf{k}\right)$ is the
amplitude of $p_{\eta}$ component with spin $\sigma$ at the $\mathrm{Cu}$($\mathrm{Ni}$)
site $\lambda$ in the band state specified by the band index $n$
and momentum $\mathbf{k}$ with energy $\epsilon_{n}\left(\mathbf{k}\right)$.
The factor $Y_{\lambda'm'\sigma',\lambda m\sigma}^{+-}\left(q\right)$
in Eq.~(\ref{eq:general}) is the density-density correlation function
of the Keldysh type, which is defined by \begin{equation}
Y_{\lambda'm'\sigma',\lambda m\sigma}^{+-}({\bf q},\omega)=\int_{-\infty}^{\infty}\langle(\rho_{{\bf q}\lambda'm'\sigma'})^{\dagger}(\tau)\rho_{{\bf q}\lambda m\sigma}(0)\rangle{\rm e}^{i\omega\tau}{\rm d}\tau,\label{eq.y+-}\end{equation}
 where \begin{equation}
\rho_{\mathbf{q}\lambda m\sigma}=\sqrt{\frac{1}{N}}\sum_{{\bf k}}d_{\mathbf{k}+\mathbf{q}\lambda m\sigma}^{\dagger}d_{\mathbf{k}\lambda m\sigma},\label{eq:rho}\end{equation}
 with \begin{equation}
d_{\mathbf{k}\lambda m\sigma}=\sqrt{\frac{1}{N}}\sum_{n}d_{n\lambda m\sigma}{\rm e}^{i\mathbf{k}\cdot\left(\mathbf{r}_{n}+\mathbf{u}_{\lambda}\right)}.\label{eq:fourier}\end{equation}
 $d_{\mathbf{k}\lambda m\sigma}$ is an annihilation operator for
the state denoted by the index $\mathbf{k}\lambda m\sigma$. The index
$\lambda m\sigma$ specifies a tight-binding orbital at the $\mathrm{Cu}$($\mathrm{Ni}$)
site $\lambda$ with $3d$ orbital $m$ and spin $\sigma$. Wavevector
$\mathbf{k}$ in eq.~(\ref{eq:rho}) runs over the first Brillouin
zone. Vectors $\mathbf{r}_{n}$ and $\mathbf{u}_{\lambda}$ in eq.~(\ref{eq:fourier})
represent a position vector of the $n$-th unit cell and the position
vectors of the $\mathrm{Cu}$($\mathrm{Ni}$) site $\lambda$ in the
unit cell $n$.

We calculate the density-density correlation function (\ref{eq.y+-})
with taking account of the correlation effect on the electron-hole
pair by the RPA as shown in fig. \ref{fig:diagram1}. For more details
of the derivation, see refs. \cite{Igarashi2006,Takahashi2007NiO,Takahashi2008JPSJ77}.

\section{Results and Discussion}

\begin{figure}
\includegraphics[scale=0.5]{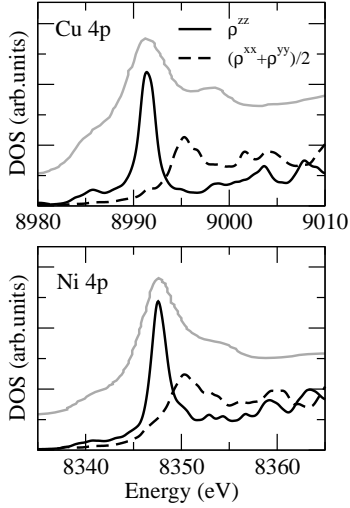}

\caption{$\rho_{\lambda4p}^{zz}$ and $\left(\rho_{\lambda4p}^{xx}+\rho_{\lambda4p}^{xx}\right)/2$
at $\mathrm{Cu}$ and $\mathrm{Ni}$ site $\lambda$ comparing with
the fluorescence yield (gray solid curve) reproduced from ref. \cite{Collart2006}.
Energy zero of the DOS is shifted so that the energy of the prominent
peak in $\rho^{zz}$ coincide with the fluorescence peak.\label{fig:4pDOS}}

\end{figure}

In order to calculate the incident-photon-dependent factor $J_{\mathrm{B}\lambda\lambda'}\left(\omega_{i},\mathbf{e}_{i};\omega_{f},\mathbf{e}_{f}\right)$,
we need the $4p$ DOS $\rho_{\lambda4p}^{\eta\eta'}(\epsilon)$ on
the $\mathrm{Cu}$($\mathrm{Ni}$) site $\lambda$. We use the $4p$
DOS given by the band structure calculation based on the LDA assuming
non magnetic states. Although the LDA fails to describe the antiferromagnetic
insulating states with a wide energy gap, the calculated $4p$ DOS
in the conduction band may be reliable because the $4p$ DOS has large
intensity in the energy range above about $10$ eV higher than the
top of the occupied states. Due to the crystal symmetry, $\rho_{\lambda4p}^{\eta\eta'}(\epsilon)$
is almost independent of $\mathrm{Cu}$($\mathrm{Ni}$) site $\lambda$,
$\rho_{\lambda4p}^{xx}(\epsilon)\approx\rho_{\lambda4p}^{yy}(\epsilon)\ne\rho_{\lambda4p}^{zz}(\epsilon)$,
and the values for $\eta\ne\eta'$ are negligible. We set the energy
difference between the $\mathrm{Cu}$($\mathrm{Ni}$) $1s$ level
and the prominent peak in $\rho_{\lambda4p}^{zz}(\epsilon)$ to be
$8992$ ($8348$) eV as shown in fig. \ref{fig:4pDOS}. Under the
condition that the dipole matrix element is constant and that the
interaction between the core hole and the $4p$ electron is neglected,
the $4p$ DOS becomes proportional to the $\mathrm{Cu}$($\mathrm{Ni}$)
$K$-edge absorption spectra. In the following, we assume that the
incident and emitted x-rays are polarized along the $z$ direction.
Accordingly, the factor $J_{\mathrm{B}\lambda\lambda'}\left(\omega_{i},\mathbf{e}_{i};\omega_{f},\mathbf{e}_{f}\right)$
is replaced with the factor $\left|L_{\mathrm{B}\lambda}^{zz}\left(\omega_{i};\omega_{f}\right)\right|^{2}$.

Another factor $Y_{\lambda'm'\sigma,\lambda m\sigma}^{+-}(q)$ mainly
determines the structure of the RIXS spectra as a function of energy
loss. We calculate this factor using the tight-binding wavefunction
given by the HFA assuming the AFM order. The $\mathrm{Cu}$($\mathrm{Ni}$)
$3d$ orbitals, apical and in-plane $\mathrm{O}$ $2p$ orbitals are
included in the tight-binding model. The Slater-Koster parameters
are taken from the LAPW band structure calculation for $\mathrm{La_{2}CuO_{4}}$\cite{DeWeert1989}.
The same values are assumed for $\mathrm{La_{2}NiO_{4}}$, because
reliable Slater-Koster parameters are not available at this stage.
Although a precise calculation based on the first principle is strongly
required, this assumption is plausible because the bond lengths $\mathrm{Cu-O}$
and $\mathrm{Ni-O}$ are nearly equal with each other\cite{Lander1989,Lander1991}
and the band structures near the Fermi level calculated within the
LDA assuming non magnetic states for $\mathrm{La_{2}CuO_{4}}$ and
$\mathrm{La_{2}NiO_{4}}$ are similar to each other\cite{Mattheiss1993}.
The intra-atomic Coulomb interaction on the $\mathrm{Cu}$($\mathrm{Ni}$)
sites is given by the Slater integrals $F^{0}$, $F^{2}$, and $F^{4}$.
Among them, $F^{2}$ and $F^{4}$, which are known to be slightly
screened by solid-state effects, are taken from the analysis of the
x-ray photoemission spectroscopy\cite{Eskes1991,Eisaki1992}. On the
other hand, $F^{0}$ is known to be considerably screened, so that
we regard the value as an adjustable parameter. The Coulomb interaction
on $\mathrm{O}$ sites and that between $\mathrm{Cu}$($\mathrm{Ni}$)
$3d$ and $\mathrm{O}$ $2p$ orbitals are absorbed into a renormalization
of the $\mathrm{O}$ $2p$ level parameters. The $\mathrm{Cu}$($\mathrm{Ni}$)
$d$-level position relative to the $\mathrm{O}$ $p$-levels is given
by the charge-transfer energy $\Delta$ defined as $\Delta=E_{d}-E_{p}+9U$
($\Delta=E_{d}-E_{p}+8U$) in the $d^{9}$ ($d^{8}$) configuration
for $\mathrm{La_{2}CuO_{4}}$ ($\mathrm{La_{2}NiO_{4}}$), where $U$
is the multiplet-averaged $d$-$d$ Coulomb interaction given by $U=F^{0}-\left(2/63\right)F^{2}-\left(2/63\right)F^{4}$\cite{Bocquet1992PRB46.3771}.
The charge-transfer energy $\Delta$ is also treated as an adjustable
parameter in the present calculation. We assume $U=11$ eV, $\Delta=2.5$
eV ($U=6$ eV, $\Delta=4$ eV) for $\mathrm{La_{2}CuO_{4}}$ ($\mathrm{La_{2}NiO_{4}}$).

\begin{figure}
\includegraphics[scale=0.45]{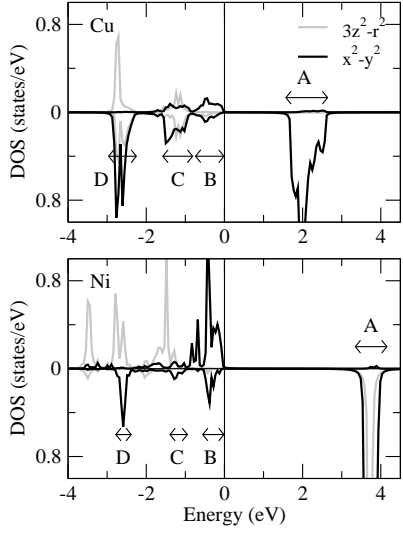}

\caption{Partial DOS of $3z^{2}-r^{2}$ and $x^{2}-y^{2}$ orbital at a transition
metal site and of $p\sigma$ orbital at an in-plane oxygen site for
$\mathrm{La_{2}CuO_{4}}$ (upper panel) and $\mathrm{La_{2}NiO_{4}}$
(lower panel). Energy zero is at the top of valence band. DOS for
the local majority and the local minority spins are presented in upper
and lower half, respectively. Roman letters A, B, C, and D are denoted
to the energy ranges where the $x^{2}-y^{2}$ partial DOS in the local
minority spin has noticeable intensity.\label{fig:pDOS}}

\end{figure}

Figure \ref{fig:pDOS} shows the DOS projected on the $\mathrm{Cu}$($\mathrm{Ni}$)
$3z^{2}-r^{2}$ and $x^{2}-y^{2}$ states. The calculated energy gap
is about $1.7$ ($3.5$) eV for $\mathrm{La_{2}CuO_{4}}$ ($\mathrm{La_{2}NiO_{4}}$).
The $x^{2}-y^{2}$ states of the local minority spin reside on both
occupied and unoccupied energy regions with noticeable weight. Therefore,
the transitions from the $x^{2}-y^{2}$ states in the energy ranges
denoted by B, C, and D to the $x^{2}-y^{2}$ states in the energy
range denoted by A contribute to the screening process of the $1s$
core-hole potential, consequently to the RIXS intensity. We note that
the transitions between the states having different local symmetry
from each other is forbidden. The distribution of the $x^{2}-y^{2}$
states of the local minority spin is similar between $\mathrm{La_{2}CuO_{4}}$
and $\mathrm{La_{2}NiO_{4}}$ except for the band width. The band
width of the states including $x^{2}-y^{2}$ state of the local minority
spin for $\mathrm{La_{2}NiO_{4}}$ is much narrower than that of $\mathrm{La_{2}CuO_{4}}$
due to the larger charge-transfer energy $\Delta$. In $\mathrm{La_{2}CuO_{4}}$,
the $3z^{2}-r^{2}$ states are completely occupied for both spins
. In $\mathrm{La_{2}NiO_{4}}$, a small amount of the $3z^{2}-r^{2}$
states of the local minority spin reside even in the occupied energy
region, while those of the local majority spin are completely occupied.
Thus, the transition in $3z^{2}-r^{2}$ channel dose not occur in
$\mathrm{La_{2}CuO_{4}}$, while the transition can contribute to
the RIXS intensity in $\mathrm{La_{2}NiO_{4}}$. In contrast to the
$e_{g}$ states, the $t_{2g}$ states cannot contribute to the RIXS
intensity, since the local majority and minority spin $t_{2g}$ states
are completely occupied.

\begin{figure}
\includegraphics[scale=0.35]{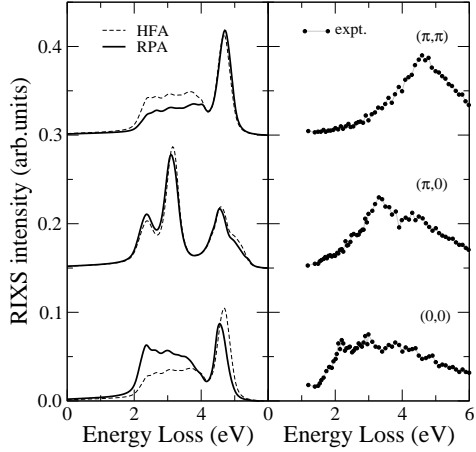}

\caption{Calculated RIXS spectra (left) and experimental spectra (right) from
ref. for $\mathrm{La_{2}CuO_{4}}$. The experimental spectra are reproduced
from ref. \cite{Ellis2008PRB77}. Solid and dotted curves are .the
spectra calculated with RPA and HFA, respectively. Incident photon
energy is assumed to be $8992$ eV, although it is $8992.5$ eV in
the experiment.\label{fig:RIXS-Cu}}

\end{figure}

Combining the two factors, we obtain the RIXS spectra with assuming
the core-hole lifetime broadening $\Gamma_{1s}=1$ eV. We convolute
the result with a Lorentzian function of $\mathrm{FWHM}=0.3\,\mathrm{eV}$
for taking account of the instrumental resolution. Figure \ref{fig:RIXS-Cu}
shows the calculated spectra for $\mathrm{La_{2}CuO_{4}}$ at several
momentum transfer $\mathbf{q}$ comparing with experiments\cite{Ellis2008PRB77}.
The RPA correction modifies the spectra calculated within HFA, leading
to better agreement with the experiments. We obtain continuous spectra
as a function of energy loss $\omega$ ranging from $\omega=2$ eV
to $6$ eV. Intensities around $\omega=2.2$, $3.2$, and $4.5$ eV
are caused by charge excitations of B$\rightarrow$A, C$\rightarrow$A,
and D$\rightarrow$A transitions, respectively, within the $x^{2}-y^{2}$
symmetry in the local minority spin states (see fig. \ref{fig:pDOS}).
A prominent peak around $4.5$ eV stays at the same position with
changing the momentum transfer $\mathbf{q}$. The spectral shape in
the low energy region changes as the momentum transfer $\mathbf{q}$
varies. A broad hump existing around $\omega=2-4$ eV at $\mathbf{q}=(0,0)$,
which is enhanced by the RPA correction, grows up to become a peak
around $3.2$ eV at $\mathbf{q}=(\pi,0)$. This behavior corresponds
well to the experimental line shape at $\mathbf{q}=(\pi,0)$, which
looks like two peaks around $\omega=3.2$ and $4.5$ eV. On the other
hand, the intensity of the broad hump around $\omega=2\sim4$ eV is
suppressed by the RPA correction at $\mathbf{q}=(\pi,\pi)$ and only
one peak is overwhelmingly left at $4.5$ eV. This explains the experimental
spectra at $\mathbf{q}=(\pi,\pi)$, which looks like a single peak
around $\omega=4.5$ eV. Although several characteristic features
of RIXS spectra are reproduced in a good agreement with the experiments\cite{Collart2006,Ellis2008PRB77},
several discrepancies between the present calculation and the experiments
still remain; the $4.5$ eV peak is hardly discernible at $\mathbf{q}=(0,0)$,
the intensity at $\omega=4$ eV is enough large forming a peak-like
structure at $\mathbf{q}=(0,0)$ with $\omega_{i}=8992$ eV\cite{Lu2006}.
To remove these discrepancies, we may need to take account of the
electron correlations beyond the RPA and the effects beyond the Born
approximation to the core-hole potential.

\begin{figure}
\includegraphics[scale=0.35]{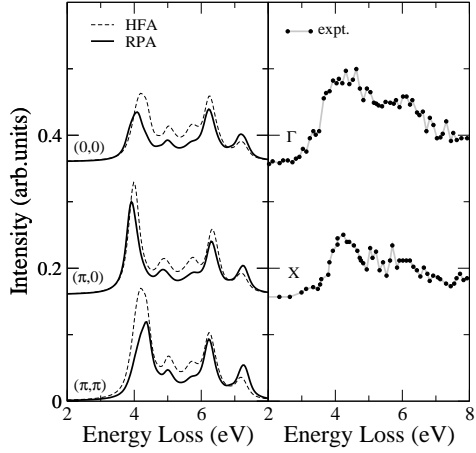}

\caption{RIXS spectra for $\mathrm{La_{2}NiO_{4}}$. The experimental spectra
are reproduced from ref. \cite{Collart2006}. Incident photon energy
is assumed to be $8348$ eV.\label{fig:RIXS-Ni}}

\end{figure}

Figure \ref{fig:RIXS-Ni} shows the calculated spectra for $\mathrm{La_{2}NiO_{4}}$
at several momentum transfer comparing with experiment\cite{Collart2006}.
We obtain continuous spectra ranging from $\omega=4$ eV to $8$ eV.
The RPA correction works to suppress the spectra calculated within
the HFA in a energy range $4\sim7$ eV at all momentum transfer $\mathbf{q}$.
The intensities around $\omega=4$, $5$, and $6.3$ eV arise mainly
from the charge excitations of B$\rightarrow$A, C$\rightarrow$A,
and D$\rightarrow$A transitions, respectively, within the $x^{2}-y^{2}$
symmetry in the local minority spin states. The transition between
the $3z^{2}-r^{2}$ states of the local minority spin also contributes
to the intensity around $\omega=5.8$ and $7.3$ eV. In contrast to
$\mathrm{La_{2}CuO_{4}}$, the calculated RIXS spectra show only slightly
peak-shifts and intensity-changes as the momentum transfer $\mathbf{q}$
changes. This non-dispersive behavior of the peak-structure as a function
of energy loss is well consistent with the experiment\cite{Collart2006}.
We note that this is a consequence of the fact that the band width
of the $x^{2}-y^{2}$ states of the local minority spin in $\mathrm{La_{2}NiO_{4}}$
is much narrower than that in $\mathrm{La_{2}CuO_{4}}$ due to the
larger charge-transfer energy $\Delta$. 

\section{Concluding Remarks}

We have analyzed the momentum transfer dependence of the RIXS spectra
in $\mathrm{La_{2}CuO_{4}}$ and $\mathrm{La_{2}NiO_{4}}$ on the
basis of the formula developed by Nomura and Igarashi. This formula
expresses the RIXS spectra by a product of the density-density correlation
function and the incident-photon dependent factor. This formula makes
it possible to calculate the RIXS spectra on the complicated models
including many orbitals, and provide clear physical interpretations
to the RIXS spectra. It also may be possible to incorporate this formula
into an ab initio band structure calculation up to the RPA level,
because the formula is similar to that of the dielectric function
$\epsilon\left(\mathbf{q},\omega\right)$. In the present study, a
multiorbital tight-binding model, which includes all the Cu(Ni) $3d$
and O $2p$ orbitals as well as the full Coulomb interaction between
$3d$ orbitals have been used to calculate the density-density correlation
function, which has been calculated from the tight-binding wavefunctions
within the HFA and RPA. The incident-photon-dependent factor has been
evaluated from the $4p$ DOS given by the ab initio band structure
calculation. 

Our results reproduced several characteristic features in the RIXS
experiments as a consequence of band-to-band transition between the
occupied and unoccupied $x^{2}-y^{2}$ states of the local minority
spin. Present calculation has also shown that in $\mathrm{La_{2}NiO_{4}}$
the RIXS peak position in terms of energy loss shifts only slightly
as the momentum transfer varies, while that looks to shift remarkably
in $\mathrm{La_{2}CuO_{4}}$, being consistent with the experiments.
This is simply explained as a consequence of the fact that the band
width of the states contributing to the RIXS intensity is significantly
narrower in $\mathrm{La_{2}NiO_{4}}$ than in $\mathrm{La_{2}CuO_{4}}$
because of a large charge-transfer energy $\Delta$. Present calculation
does not reproduce all features at this stage partly because of the
strong correlation between $3d$ electrons. In spite of several drawbacks
of our analysis, it is suggested that the peak shifts as a function
of momentum should not be interpreted as a dispersion relation of
a kind of exciton, because the peaks with broad widths constitute
an energy continuum generated by a band-to-band transition. We have
shown that the weight of the $x^{2}-y^{2}$ states of the local minority
spin in the occupied band states strongly depends on the momentum\cite{Takahashi2008JPSJ77}.
The momentum dependence of the weight of the $x^{2}-y^{2}$ states
of the local minority spin in the band states and the dispersion relation
of the band states mainly determine the momentum transfer dependence
of the RIXS intensity in the cuprate and nickelate. 

Although experimental data have been accumulated for the doped cuprates
and nickelates\cite{Ishii2005-1,Ishii2005-2,Lu2005,Collart2006},
theoretical analyses are limited on a one-band Hubbard model within
the exact diagonalization method\cite{Tsutsui2003}, and on a three-band
Hubbard model analysis within the HFA on the basis of the present
formalism\cite{Markiewicz2006}. An analysis with a detailed model
like the present paper may be necessary to clarify the momentum and
incident-photon dependence of the spectra. Since electron correlations
are expected to be more important in the doped cuprates and nickelates,
such studies seem rather hard and are left in future.

\ack{}{This work was partly supported by Grant-in-Aid for Scientific Research
from the Ministry of Education, Culture, Sport, Science, and Technology,
Japan.}

\section*{References}

\bibliography{BibQSD2008}

\begin{thebibliography}{10}

\bibitem{Kim2002}
Y.~J. Kim, J.~P. Hill, C.~A. Burns, S.~Wakimoto, R.~J. Birgeneau, D.~Casa,
  T.~Gog, and C.~T. Venkataraman.
\newblock 2002, {\em Phys. Rev. Lett.}, {\bf 89} 177003.

\bibitem{Lu2006}
L.~Lu, J.~N. Hancock, G.~Chabot-Couture, K.~Ishii, O.~P. Vajk, G.~Yu,
  J.~Mizuki, D.~Casa, T.~Gog, and M.~Greven.
\newblock 2006, {\em Phys. Rev. B}, {\bf 74} 224509.

\bibitem{Ellis2008PRB77}

\newblock D.~S. {Ellis}, J.~P. {Hill}, S.~{Wakimoto}, R.~J. {Birgeneau},
  D.~{Casa}, T.~{Gog}, and Y.-J. {Kim} 2008, {\em Phys. Rev. B}, {\bf 77}
  060501.

\bibitem{Collart2006}
E.~Collart, Abhay Shukla, J.-P. Rueff, P.~Leininger, H.~Ishii, I.~Jarrige,
  Y.~Q. Cai, S.-W. Cheong, and G.~Dhalenne.
\newblock 2006, {\em Phys. Rev. Lett.}, {\bf 96} 157004.

\bibitem{Zhang1998PRB58}

\newblock F.~C. Zhang and K.~K. Ng 1998, {\em Phys. Rev. B}, {\bf 58} 13520.

\bibitem{Nomura2004}
Takuji Nomura and Jun-ichi Igarashi.
\newblock 2004, {\em J. Phys. Soc. Jpn.}, {\bf 73} 1677.

\bibitem{Nomura2005}

\newblock Takuji Nomura and Jun-ichi Igarashi 2005, {\em Phys. Rev. B}, {\bf
  71} 035110.

\bibitem{Nozieres1974}

\newblock Philippe Nozi\`eres and Elihu Abrahams 1974, {\em Phys. Rev. B}, {\bf
  10} 3099.

\bibitem{Suga2005}

\newblock S.~Suga, S.~Imada, A.~Higashiya, A.~Shigemoto, S.~Kasai, M.~Sing,
  H.~Fujiwara, A.~Sekiyama, A.~Yamasaki, C.~Kim, T.~Nomura, J.~Igarashi,
  M.~Yabashi, and T.~Ishikawa 2005, {\em Phys. Rev. B}, {\bf 72} 081101.

\bibitem{Igarashi2006}

\newblock Jun-ichi Igarashi, Takuji Nomura, and Manabu Takahashi 2006, {\em
  Phys. Rev. B}, {\bf 74} 245122.

\bibitem{Takahashi2008JPSJ77}
M~Takahashi, J~Igarashi, and T~Nomura.
\newblock 2008, {\em J. Phys. Soc. Jpn.}, {\bf 77} 034711.

\bibitem{Takahashi2007NiO}
M~Takahashi, Junichi Igarashi, and T~Nomura.
\newblock 2007, {\em Phys. Rev. B}, {\bf 75} 235113.

\bibitem{Nozieres1969_1097}
P.~Nozi{\`e}res and C.~T.~De Dominicis.
\newblock 1969, {\em Phys. Rev.}, {\bf 178} 1097.

\bibitem{DeWeert1989}
M.~J. DeWeert, D.~A. Papaconstantopoulos, and W.~E. Pickett.
\newblock 1989, {\em Phys. Rev. B}, {\bf 39} 4235.

\bibitem{Lander1989}

\newblock G.~H. Lander, P.~J. Brown, J.~Spal/ek, and J.~M. Honig 1989, {\em
  Phys. Rev. B}, {\bf 40} 4463.

\bibitem{Lander1991}

\newblock G.~H. Lander, P.~J. Brown, C.~Stassis, P.~Gopalan, J.~Spalek, and
  G.~Honig 1991, {\em Phys. Rev. B}, {\bf 43} 448.

\bibitem{Mattheiss1993}

\newblock L.~F. Mattheiss 1993, {\em Phys. Rev. B}, {\bf 48} 4352.

\bibitem{Eskes1991}
H.~Eskes and G.~A. Sawatzky.
\newblock 1991, {\em Phys. Rev. B}, {\bf 43} 119.

\bibitem{Eisaki1992}

\newblock H.~Eisaki, S.~Uchida, T.~Mizokawa, H.~Namatame, A.~Fujimori, J.~van
  Elp, P.~Kuiper, G.~A. Sawatzky, S.~Hosoya, and H.~Katayama-Yoshida 1992, {\em
  Phys. Rev. B}, {\bf 45} 12513.

\bibitem{Bocquet1992PRB46.3771}

\newblock A.~E. Bocquet, T.~Mizokawa, T.~Saitoh, H.~Namatame, and A.~Fujimori
  1992, {\em Phys. Rev. B}, {\bf 46} 3771.

\bibitem{Ishii2005-1}
K.~Ishii, K.~Tsutsui, Y.~Endoh, T.~Tohyama, K.~Kuzushita, T.~Inami, K.~Ohwada,
  S.~Maekawa, T.~Masui, S.~Tajima, Y.~Murakami, and J.~Mizuki.
\newblock 2005, {\em Phys. Rev. Lett.}, {\bf 94} 187002.

\bibitem{Ishii2005-2}
K.~Ishii, K.~Tsutsui, Y.~Endoh, T.~Tohyama, S.~Maekawa, M.~Hoesch,
  K.~Kuzushita, M.~Tsubota, T.~Inami, J.~Mizuki, Y.~Murakami, and K.~Yamada.
\newblock 2005, {\em Phys. Rev. Lett.}, {\bf 94} 207003.

\bibitem{Lu2005}
L.~Lu, G.~Chabot-Couture, X.~Zhao, J.~Hancock, N.~Kaneko, O.~Vajk, G.~Yu,
  S.~Grenier, Y.-J. Kim, D.~Casa, and M.~Greven.
\newblock 2005, {\em Phys. Rev. Lett.}, {\bf 95} 217003.

\bibitem{Tsutsui2003}

\newblock Kenji Tsutsui, Takami Tohyama, and Sadamichi Maekawa 2003, {\em Phys.
  Rev. Lett.}, {\bf 91} 117001.

\bibitem{Markiewicz2006}
R.~S. Markiewicz and A.~Bansil.
\newblock 2006, {\em Phys. Rev. Lett.}, {\bf 96} 107005.

\end{thebibliography}

\end{document}